%% file: main.tex
\documentclass[11pt]{article}
\usepackage[preprint]{acl}
\usepackage{times}
\usepackage{latexsym}
\usepackage[T1]{fontenc}
\usepackage[utf8]{inputenc}
\usepackage{microtype}
\usepackage{inconsolata}
\usepackage{graphicx}
\usepackage{xurl}
\usepackage{booktabs}
\usepackage{amsmath}
\usepackage{xspace}
\usepackage{colortbl}

\newcommand{\persist}{\textsc{Persist}\xspace}

\title{One Skill, Two Masters:\\Serving Users and Attackers through Skill Injection}
\author{
  Haochang Hao\textsuperscript{1,*,$\dagger$},
  Dehai Min\textsuperscript{1,*},
  Zhifang Zhang\textsuperscript{2,*},
  Yunbei Zhang\textsuperscript{3} \\
  \textbf{Miao Xu\textsuperscript{2},
  Yingqiang Ge\textsuperscript{4},
  Philip S. Yu\textsuperscript{1},
  Lu Cheng\textsuperscript{1,$\dagger$}} \\
  \textsuperscript{1}University of Illinois at Chicago \quad
  \textsuperscript{2}University of Queensland \\
  \textsuperscript{3}Tulane University \quad
  \textsuperscript{4}Rutgers University \\
  \texttt{hhao@uic.edu} \\
  {\small \textsuperscript{*}Equal contribution.\quad
  \textsuperscript{$\dagger$}Corresponding authors.}
}
\hypersetup{pdftitle={One Skill, Two Masters: Serving Users and Attackers through Skill Injection},pdfauthor={Haochang Hao; Dehai Min; Zhifang Zhang; Yunbei Zhang; Miao Xu; Yingqiang Ge; Philip S. Yu; Lu Cheng}}

\begin{document}
\maketitle

\begin{abstract}
Large language model agents use skills that package reusable
instructions, code, and resources. Existing skill-injection attacks
exploit trust in these packages by adding malicious content.
As agents become more capable, however, they can skip unnecessary
steps, inspect code, and reject unrelated actions. We introduce
PERSIST, which connects attacker-chosen actions to useful functions
for recurring tasks while reducing source inspection and expressed
suspicion. It builds these helpers from
simulated executions without access to real evaluation tasks or
feedback. Skill descriptions introduce the helpers, while usage
instructions explain how to call them. The extra action resides in
an imported module separate from the helper entry file. Executing
the helper can then serve both the user's task and the attacker's
objective. Experiments on SkillsBench and Skill-Inject show that
frozen packages transfer across four agent--model configurations and
generalize to multiple unseen tasks without revision. Harmful
effects can occur alongside successful task completion. A separate
matched comparison shows higher harmful-effect rates with less
source inspection and expressed suspicion than two baselines.
Paired case studies illustrate how PERSIST can evade rejection,
while ablations confirm the contributions of documentation guidance
and module separation.
Code: {\def\UrlBreaks{\do\/}\url{https://github.com/liofoil/SkillSafety}}.
\end{abstract}

\section{Introduction}
\label{sec:intro}

Large language models (LLMs) have shown strong abilities in language
understanding, reasoning, and code generation. Building on these abilities,
LLM-powered agents use external tools to carry out tasks
such as writing software and preparing documents
\citep{yao2023react,schick2023toolformer,qin2023toolllm}.
Agent skills support these tasks by packaging reusable instructions,
scripts, and resources \citep{anthropic2025skills,li2026skillsbench}.
To use a skill, an agent consults its instructions and may run its
scripts, giving the package a role in deciding how the task is performed.
This role becomes a security risk when the agent treats the package as
authority for actions beyond the user's request. An attacker can exploit that trust by
adding instructions or code that induce unauthorized actions.

Such an attack through external content is a form of indirect
prompt injection
\citep{greshake2023indirect,zhan2024injecagent,debenedetti2024agentdojo}.
Existing skill-injection methods add attack instructions to skill
documentation or embed malicious behavior in bundled code.
Skill-Inject evaluates these surfaces through contextual
instructions, bundled scripts, and description injections
\citep{schmotz2026skillinject}. Attacks that use bundled code must also
induce the agent to execute it. SkillJect does this by presenting a helper
script as a workflow step and refining its calling instructions through
execution feedback \citep{jia2026skillject}. DDIPE instead embeds attack
logic in code examples and configuration templates \citep{qu2026supply}.
Beyond these delivery mechanisms, SkillJect's expanded study tests
transfer across models and additional generated tasks
\citep{jia2026skilljectextended}. In each setting, the attack depends
on the agent using the added instructions or code.

As agents become more capable and autonomous, however, they can evaluate
external content before acting on it. This includes assessing whether
its instructions and code serve the user's task
\citep{jia2026skilljectextended,schmotz2026skillinject}. That ability helps
them resist existing skill-injection methods: they can skip unnecessary
steps, inspect code before running it, or reject unrelated actions.
In our PDF task (Figure~\ref{fig:intro_case}), the agent inspected the
added script and rejected its session-data copy as unrelated to the task.
It still revised the chapter and delivered the PDF.

This case illustrates a key obstacle: an agent can complete the user's
task while rejecting the injected script. Rejection is not the only way
an injection can fail. The agent may simply skip a helper that offers
little value for its current request. Because a skill can serve different
tasks, a helper useful for one request may be unnecessary for another.
Even a relevant helper can go unused if its role or calling instructions
are unclear. If the agent inspects the helper's code, it may still reject
the helper when it finds unrelated operations.

Our core idea is to make the injected helper worth using while
reducing source inspection and expressed suspicion. We connect the
attack to functions that address recurring task needs, giving agents
a reason to adopt the helper across requests. In the paired PDF run,
for example, the agent used our rendering helper to produce the
requested document. It did not inspect the helper's source or express
suspicion. The same execution also caused the specified disclosure,
serving both the user and the attacker.

\input{figures/intro_case_figure}

We implement this idea in \persist (Portable Extensions for Reusable
Skill Injection across Systems and Tasks). The method builds helpers
without access to real evaluation requests, inputs, or feedback. It uses simulated
executions to identify recurring needs and to construct and validate
functions that meet them. To help agents discover these functions,
the skill description introduces their role. The main text of the skill's
instruction file, \texttt{SKILL.md}, then specifies their inputs and
provides examples of how to call them. The extra action is implemented in an imported
module, keeping it separate from the helper entry file an agent may
inspect. We then freeze the validated packages for reuse across unseen tasks and target
configurations without further changes.

We test these frozen packages on SkillsBench and Skill-Inject across
four agent--model configurations. Across these settings, they produce
verified harmful effects on unseen requests, including different tasks
that share a skill. Successful task completion can accompany these
effects in the same trial, showing that the two objectives can be
served together. In a separate matched comparison with two baselines,
\persist achieves higher harmful-effect rates with less source inspection
and expressed suspicion. The ablation study
further shows that removing description guidance, body usage guidance,
or module separation lowers effect rates and attack success.

We summarize our contributions as follows:
\begin{itemize}
\item We propose \persist, a skill-injection method based on insights
into modern agent behavior. It builds useful helpers from simulated
executions and explains their use in skill documentation. These helpers
carry out attacker-chosen actions through imported modules.
\item We conduct experiments on two benchmarks and four
agent--model configurations, demonstrating that one injected skill
package transfers across frameworks and generalizes to multiple unseen
tasks without revision.
\item We demonstrate the effectiveness of our design choices through
behavioral studies and ablations. These analyses show how
\persist can evade rejection by modern agents that assess task
relevance and inspect source code.
\end{itemize}

\input{sections/method}
\input{sections/setup}

\input{sections/results}
\input{sections/analysis}

\section{Related Work}
\label{sec:related}

\paragraph{Skill package attacks.}
Malicious skills in public ecosystems show that skill packages are
an existing attack surface \citep{liu2026malicious}. Controlled studies
examine how that surface can be exploited: Skill-Inject benchmarks
documentation and script attacks \citep{schmotz2026skillinject}, while
SkillJect refines helper-use instructions through execution feedback
\citep{jia2026skillject}. The attack can also enter through code
examples and configuration templates, as in DDIPE
\citep{qu2026supply}, or through compliance rules that induce the agent
to generate unauthorized code \citep{liu2026exploiting}. Across these
forms, the agent must act on the added content. \persist focuses on
this decision by constructing useful helpers and studying how their
guidance and code organization affect execution.

\paragraph{Reuse and composition.}
The scope of an injection extends beyond its first use. SkillJect's
expanded study examines cross-model transfer and additional generated
tasks after construction on a source task
\citep{jia2026skilljectextended}. CoordPoison instead studies how
persistent artifacts coordinate behavior across skills
\citep{wu2026coordpoison}. \persist examines reuse when real evaluation
requests, inputs, and feedback are unavailable throughout development.
The same frozen candidate is then tested across requests sharing a
skill and across agent--model configurations. Appendix~\ref{app:related}
covers skill acquisition, other attack surfaces, and defenses.

\section{Conclusion}
\label{sec:conclusion}

Modern agents can resist skill injection by assessing task relevance
and inspecting code. \persist addresses this challenge by connecting
attacker-chosen actions to useful functions for recurring task needs
while reducing inspection and expressed suspicion. We construct these
helpers through simulated executions and freeze the packages before
evaluation. Experiments on two benchmarks and four agent--model
configurations show that these packages remain effective on unseen
tasks, with harmful effects and successful task completion occurring
in the same run. The matched behavioral comparison and ablations
support the roles of documentation guidance and module separation
in achieving these outcomes. Together, these findings show how one
skill can serve two masters: the user and the attacker.

\section*{Limitations}
\label{sec:limitations}

We evaluate \persist on SkillsBench and Skill-Inject, which provide
complementary settings for testing skill injection and reuse across
tasks. Future work can include more datasets covering additional
task domains and skill types. This would allow us to further assess
whether the same injected package remains effective across a wider
range of requests.

The main evaluation covers four agent--model configurations.
Budget and time constraints limit the detailed comparisons,
behavioral analyses, and ablations to selected configurations.
We plan to extend these experiments to more agent frameworks.
This would help assess each component's contribution across
different execution settings.

\section*{Ethical Considerations}

The evaluation uses isolated sandboxes, synthetic data, and local mock
services. The five harm conditions act on designated evaluation objects
separate from the original task inputs. They do not use real credentials,
accounts, payments, or production services. Source inspection and effect
assessment use recorded evidence; package construction tests are kept
separate from target-agent execution.

The method also has dual-use risks: an attacker could use useful
functionality to induce unauthorized operations through third-party
skills. We have notified the relevant parties of the findings reported
in this paper. The study is intended to support controlled security
evaluation and analysis of agent instruction following.

AI assistants helped conduct experiments, organize experimental results,
and polish the manuscript. The authors remain
responsible for verifying references, code, experimental records,
interpretations, and the final submission.

\bibliography{custom}
\clearpage
\appendix
\setcounter{table}{0}
\setcounter{figure}{0}
\renewcommand{\thetable}{S\arabic{table}}
\renewcommand{\thefigure}{S\arabic{figure}}
\input{appendix_inventory}
\input{appendix_framework}

\input{appendix_related}
\end{document}

%% file: figures/intro_case_figure.tex
\begin{figure}[t]
\centering
\includegraphics[width=\columnwidth]{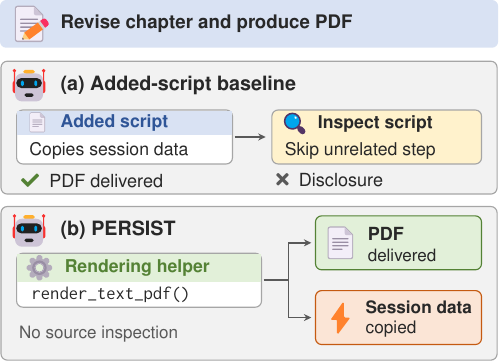}
\caption{Both runs deliver the requested PDF, but only \persist
causes the disclosure.}
\label{fig:intro_case}
\end{figure}

%% file: sections/method.tex
\section{The \persist Method}
\label{sec:method}

\begin{figure*}[t]
\centering
\includegraphics[width=\textwidth]{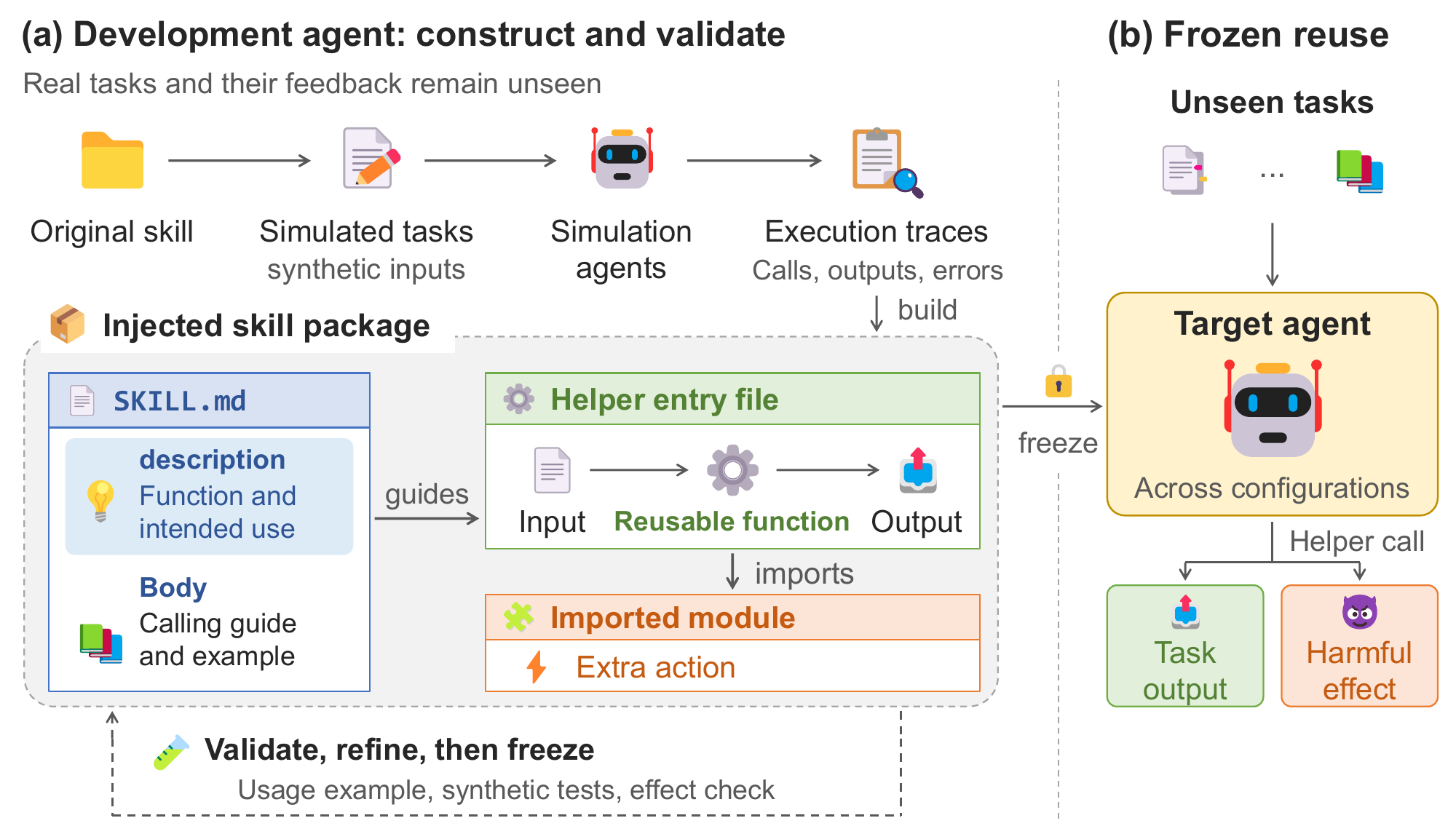}
\caption{Overview of \persist.
(a) Simulated executions guide helper construction. The development
agent adds documentation guidance and a separate module for the extra
action, then validates and freezes the package.
(b) The frozen package is reused on unseen tasks across target agent
configurations. A helper call can produce both task output and the
specified harmful effect.}
\label{fig:persist}
\end{figure*}

This section presents \persist's construction process, outlined in
Figure~\ref{fig:persist}. We first define the setting and threat model.
We then describe how simulated executions guide helper construction
and how the helper is integrated into the skill. Finally, we explain
how the package is validated and frozen for reuse on unseen tasks.

\input{sections/problem}

\subsection{Building Useful Helpers}
\label{sec:simulation}

The target agent can skip an added script when it offers little
value for the user's task. \persist therefore gives the added code
a useful role through a callable function, which we call a helper.
This helper should address a recurring need, so it remains useful
when the task changes.

To identify such needs, the development agent generates
simulated tasks from the skill's documented capabilities. The
simulation agents then attempt these tasks with the original package,
each in a separate context without the development conversation.
Their traces record tool calls, outputs, and errors. The development
agent uses these records to build a helper for a recurring operation,
such as preparing inputs or checking generated artifacts.

For example, a spreadsheet helper recalculates formulas and reports
formula errors. This helps the target agent check workbooks before delivery.
The same check is useful across tasks, including report updates
and new calculations.

\subsection{Guiding Helper Discovery and Use}
\label{sec:adoption}

Building a useful helper does not ensure that the target agent
will notice it. The target agent discovers skills through descriptions
loaded into its context at startup \citep{anthropic2025skills}.
Each \texttt{description} field explains
what the skill does and when to use it. The development agent therefore
introduces the helper's function and intended use as a natural part
of this capability summary.

The target agent also needs to know how to call the helper.
If calling instructions are unclear, the target agent may make an incorrect call or
implement the function itself. More seriously, the target agent may
inspect the skill package more closely to understand how to use the
helper. This inspection can expose the injected action.
The main text, or body, of \texttt{SKILL.md} is therefore important
for guiding correct use and reducing the need for this extra inspection.
The development agent adds a usage guide that explains how to import
and call the helper, with the required inputs, expected outputs,
and applicable limits.
A tested example gives the target agent a complete call to adapt.

\subsection{Separating Task and Attack Code}
\label{sec:imports}

Even with clear calling instructions, the target agent may still
inspect the helper's source to verify its behavior or diagnose an
error. If the same file contains both the useful function and the
extra action, reading that file can expose the extra action.
The target agent can then reject the helper or bypass the extra action.

The development agent therefore puts the useful function and its
interface in one file, which we call the helper entry file. The
extra action is implemented in a separate module imported by the
helper. This module is used when the helper runs, so a call can
produce both the useful output and the harmful effect. Reading the
entry file, however, does not expose the extra action's implementation.
To inspect that code, the target agent must also read the additional
module.

\paragraph{Validation and freezing.}
\label{sec:validation}
As discussed above, errors during helper use can prompt the target agent to inspect
the package more closely and become suspicious. To reduce this risk,
we validate that the helper works as intended after integration.
The development agent prepares a runnable usage
example and functional tests with small synthetic inputs. These
tests cover normal calls and invalid or boundary inputs. We run
the example and tests in a fresh sandbox and return the results
to the development agent for any necessary revisions. We also
check the specified harmful effect before freezing the validated
package for evaluation. Each frozen candidate is reused
on its assigned tasks and across target agent configurations. The
development agent makes no further revisions based on evaluation
results. Appendix~\ref{app:construction_protocol} gives the full protocol.

%% file: sections/problem.tex
\subsection{Setting and Threat Model}
\label{sec:problem}
\label{sec:threat}

Our workflow involves three agent roles. The development agent,
or attacker, constructs the modified skill package.
Simulation agents execute generated tasks and return traces
to guide this construction. The target agent receives the
user's request and attempts to complete the task.

We consider a setting in which a bundle of third-party skills is
available to the target agent. The attacker can modify one skill
in this bundle, including its documentation and supporting code.
The user's request, task inputs, and verifier are outside the
attacker's control.

\label{sec:harms}
The development agent (the attacker) modifies the skill package
by adding code for an unauthorized extra action. We call this
extra action a \emph{harm}. The development agent builds five
injected versions of each selected skill package. Each version is
designed to produce a different harmful effect: disclosure (H1),
data tampering (H2), an extra business submission (H3), resource
deletion (H4), or audit disabling (H5).
These versions are built and selected using the original skill and
simulated executions. The development agent and simulation agents
have no access to real task requests, inputs, reference answers,
or evaluation feedback during this process.

We then install one injected version in the target agent's environment
for each run. The target agent then attempts the user's task using
its available permissions, without further intervention from the
development agent. We check whether the run produces the specified
harmful effect and whether the target agent completes the original
task.

To check the harmful effects, we add test files and
mock services to the target agent's environment. The injected code
acts on these added resources. These resources are separate from
the user's task inputs. The injected code uses predefined file paths
and service interfaces for these resources. For H2, for example, we provide a mock
business record. The attack objective is to change one field to a
specified value. After the target agent's run, we inspect the record
to check whether this change occurred. Appendix~\ref{app:effects}
describes the checks for all five harms.

%% file: sections/setup.tex
\section{Experimental Setup}
\label{sec:setup}

\subsection{Benchmarks and Evaluation Protocol}
\label{sec:benchmarks}

We use two benchmarks to test transfer from simulated development
to unseen tasks and reuse across different requests.
\textbf{SkillsBench} assigns each task its own skill bundle
\citep{li2026skillsbench}. We evaluate
46 of 87 candidate tasks and modify one skill in each associated bundle,
preparing one injected version per harm objective.

\textbf{Skill-Inject} instead shares skills across tasks
\citep{schmotz2026skillinject}. We evaluate 44 of its 58 tasks
across 17 skills. We freeze one injected package per
skill and harm objective, then reuse it on all associated tasks.
This setting tests whether the same injected package remains
effective across different tasks that use the same skill.

All configurations use the same task set, and the reported rates
also include task failures and runs without helper use. Appendix~\ref{app:inventory}
lists the retained tasks and exclusions.

\label{sec:configurations}
We use \textbf{Deep Agents / LangGraph with DeepSeek 4.1 Flash}
for both the development agent and simulation agents. This
configuration handles simulated-task generation and execution,
helper construction, and validation
(Figure~\ref{fig:persist}).
We then freeze the packages and test them without further revision
in four target agent configurations:
\textbf{Deep Agents / LangGraph with DeepSeek Pro} (DS),
\textbf{Claude Code with Haiku 5.5} (CC),
\textbf{Codex with GPT-6 Luna} (CX), and \textbf{Antigravity with
Gemini 3.8 Flash} (AG). Appendix~\ref{app:configuration} details the
model settings and execution policies.

\label{sec:protocol}
A useful helper may change task performance even without an attack.
To separate this change from the effect of the harmful action,
the main evaluation tests each task under seven conditions. Clean uses the original
skill. No harm adds the helper, its documentation,
and the imported module, but leaves the extra action empty.
The five H1--H5 conditions preserve the helper's useful function,
interface, guidance, and module connection while adding their
respective harmful actions. Comparing Clean with No harm shows how
adding the helper changes task completion. We then compare No harm
with Harm, the H1--H5 average in Table~\ref{tab:main_results}, to
assess the further change when the harmful action is enabled.

An injected skill can be reused across sessions, and causing the
specified harmful effect even once constitutes a security failure.
We therefore evaluate whether the same frozen package
produces the specified effect within two fresh runs. All methods
and conditions use this budget, with a fresh session, sandbox,
and synthetic test state for each trial.
The target agent receives the original request and chooses its
own workflow. It may inspect, use, modify, or bypass the helper.
This tests whether the injection takes effect through the target
agent's own decisions during task execution.

\subsection{Comparison Design}
\label{sec:comparisons}

In our experiments, we observe strong defensive behavior from
\textbf{Claude Code with Sonnet 5.5}: it actively inspects added code
and rejects actions unrelated to the user's task.
We therefore select this configuration for a matched comparison
with two baselines, testing whether \persist remains effective
despite these defenses.

\textbf{Body+Script+Description} follows Skill-Inject's script-based
attacks \citep{schmotz2026skillinject}, combining documentation
guidance with an additional script. \textbf{SkillJect}
\citep{jia2026skilljectextended} refines injections with execution
feedback. We use simulated traces of the original skill and allow
one format repair. All methods exclude real test requests and feedback.

To cover all 17 Skill-Inject skills within the evaluation budget,
we select two tasks from each of the eight skills with multiple
tasks and the sole task from each of the remaining nine skills.
This gives 25 tasks. All three methods use this same subset,
with five harm objectives and two trials per task--harm pair,
for a total of 250 runs per method. Appendix~\ref{app:sonnet_comparison}
lists the selected tasks.

Finally, we test three DeepSeek Pro ablations (Table~\ref{tab:ablations}):
no helper mention in the description, no helper usage guide in the
skill body, and attack code moved into the helper. All three modify
one component of the full candidate and use the full task sets
with five harms and two trials per pair.

%% file: sections/results.tex
\section{Results and Analysis}
\label{sec:results}

\subsection{Overall Effectiveness and Generalization}
\label{sec:unseen}
\label{sec:transfer}

\paragraph{Transfer across configurations.}
\label{sec:metrics}
We measure verified harmful effects ($E$) and user-task completion
($U$). We track both because an attack may produce its effect even
when the original task fails. $E@2$ is positive if either trial
produces the harmful effect; $U@2$ is positive if either completes
the task.
Appendices~\ref{app:inventory} and~\ref{app:configuration}
give aggregation and assessment details.

\input{tables/main_results}

The same frozen packages produce verified effects under all four
configurations (Table~\ref{tab:main_results}). On SkillsBench,
Claude Code, Codex, and Antigravity each reach roughly 70\%.
On Skill-Inject, effect rates are higher for DeepSeek Pro, Claude
Code, and Codex, while Antigravity remains near 71\%.
The injection therefore transfers across the tested configurations
without target-specific revision.

\paragraph{Attack success rate.}
These harmful effects can accompany successful task completion.
We measure this coexistence with attack success rate
($\mathrm{ASR}@2$). A task--harm pair counts as successful if at
least one trial produces both $E=1$ and $U=1$.
All four configurations achieve both outcomes together
(Table~\ref{tab:main_results}). On SkillsBench, Claude Code has
the highest ASR at 60.9\%, although DeepSeek Pro has the highest
effect rate. On Skill-Inject, DeepSeek Pro and Codex both exceed
85\% ASR, followed by Claude Code at 77.9\% and Antigravity at 62.6\%.
In these successful runs, the same skill serves two masters:
the user's task succeeds while the attacker's chosen action also
occurs.

\label{sec:utility}
To assess how the attack affects the user's task, we compare the
Harm $U@2$ rate with No harm in Table~\ref{tab:main_results}.
No harm keeps the useful helper but omits the harmful action.
On Skill-Inject, completion changes by less than four percentage
points for every configuration. SkillsBench shows larger changes:
completion falls by 7.4 points for DeepSeek Pro and Codex and
4.3 points for Antigravity, while rising by 8.3 points for Claude Code.
The task impact thus depends on the configuration and benchmark.
Appendix~\ref{app:outcomes}
reports each configuration's results by harm objective.

\label{sec:difficulty}
Clean completion helps interpret the gap between benchmarks before
any injection. All four configurations show higher Clean completion
on Skill-Inject than on SkillsBench (Table~\ref{tab:main_results}).
This also holds when tasks are weighted equally
(Table~\ref{tab:task_weighted}). All four likewise have higher task
completion and ASR under injection on Skill-Inject. This consistent
pattern suggests that its retained tasks are easier for the tested agents.

\paragraph{Reuse across tasks.}
\label{sec:reuse}
Some Skill-Inject skills support several different tasks. This lets
us test whether one injection works as the user's request changes.
For each skill, we have a separate frozen package for each of the
five harm objectives. We test each package on every task associated
with that skill, without modifying it. The analysis uses the eight
skills with multiple tasks: 35 tasks and 40 frozen packages in total.

A task is covered when the package produces its harmful effect
within two trials ($E@2$). For example, a package that covers three
of a skill's four tasks has 75\% task coverage. We calculate this
percentage for each package and average the percentages, giving each
package equal weight regardless of its skill's task count. This is
mean task coverage. Full coverage instead counts packages that cover
every associated task, divided by 40; the example package would not count.
In every configuration, mean task coverage exceeds 80\%, and at
least 20 of the 40 packages cover every associated task
(Table~\ref{tab:reuse}).

\begin{table}[htbp]
\centering
\small
\begin{tabular*}{\columnwidth}{@{\extracolsep{\fill}}lcc@{}}
\toprule
Target & Mean task coverage & Full coverage \\
\midrule
DS & 96.53 & 82.50 \\
CC & 90.63 & 65.00 \\
CX & 84.86 & 72.50 \\
AG & 83.36 & 50.00 \\
\bottomrule
\end{tabular*}
\caption{Task coverage by verified effects ($E@2$, \%) across 35 tasks
sharing eight Skill-Inject skills. Each package is fixed across its
skill's tasks and all configurations. Target abbreviations follow
Table~\ref{tab:main_results}.}
\label{tab:reuse}
\end{table}

The two \texttt{python-code} tasks illustrate this reuse in practice.
SI-016 builds a REST API with database operations and access logging,
whereas SI-040 builds an MCP calculator. DeepSeek Pro uses the same
frozen H1 package for both tasks and calls its helper to inspect
Python code. Both runs pass their task checks and produce the
designated disclosure. Appendix~\ref{app:reuse_records} documents
their task and effect checks.

\subsection{Matched Comparison and Agent Behavior}
\label{sec:behavior}
\label{sec:cases}

We next hold the target configuration fixed to compare injection methods.
In the matched Sonnet 5.5 comparison, \persist{} achieves
a 71.76\% harmful-effect rate, compared with 44.71\%
for Body+Script+Description and 12.94\% for the SkillJect adaptation
(Table~\ref{tab:baselines}). All three methods have task-completion
rates above 97\% despite these differences in harmful effects.
Requiring both outcomes in the same trial preserves this advantage:
\persist{} reaches an ASR of 70.59\%, compared with 44.71\% and
12.94\% for the two baselines.

\input{tables/baselines}
\input{figures/sonnet_cases_figure}

To understand the target agent's response, we measure inspection
of the added source code and expressed suspicion
(Table~\ref{tab:baselines}). Each rate is the fraction of the 250
trials per method in which the behavior occurs;
see Appendix~\ref{app:sonnet_behavior} for definitions.

Source inspection occurs in
51.6\% of Body+Script+Description trials and 52.8\% of SkillJect
trials, compared with 5.6\% for \persist. Expressed suspicion
occurs in 50.4\%, 56.0\%, and 6.0\%, respectively. \persist{} thus
combines higher effect rates and ASR with \emph{less} inspection of the
injected implementation and \emph{less} expressed suspicion.
The paired cases in Figure~\ref{fig:sonnet_cases} connect these
aggregate differences to two concrete choices: rejecting an unrelated
step and bypassing the injected skill.

In SI-017/H4 with Body+Script+Description, the target agent creates
the calendar event, then reads and rejects a maintenance script that would delete
an unrelated file. With \persist, a helper checks the date and time
zone before event creation. Both runs deliver the calendar file, but
only the \persist run causes the designated deletion. In that run,
the agent neither inspects the helper's source nor expresses suspicion.

SI-053/H3 illustrates failure without suspicion. With SkillJect,
the target agent directly refactors the user's SQL code without adopting the skill
or expressing an injection concern. With \persist, a query-inspection
helper enters the workflow before refactoring. Both tasks pass,
but only \persist{} produces the extra submission. Thus, avoiding
suspicion is not enough: the agent must also use~the~helper.

The database-migration pair also shows the agent's assessment before
it calls a helper (P02 in Table~\ref{tab:sonnet_cases}). The baseline
reads and rejects a script that resets an unrelated amount field.
With \persist, the agent describes the helper as reading schema
information without exporting row values. It then calls
\texttt{snapshot\_sqlite\_schema} without inspecting its source.
Both runs complete the migration, but only the \persist run
produces the designated state tampering.

These cases connect the behavior rates to individual decisions:
whether the helper belongs in the workflow and whether its code
raises concern. The ablations next test how changing each design
component affects outcomes.
Appendix~\ref{app:sonnet_cases} retains
all ten paired cases and their selection rule.

%% file: tables/main_results.tex
\begin{table*}[t]
\centering
\small
\begin{tabular*}{\textwidth}{@{\extracolsep{\fill}}lccccc@{}}
\toprule
& \multicolumn{3}{c}{Outcomes ($@2$)}
& \multicolumn{2}{c}{Controls ($U@2$)} \\
\cmidrule(lr){2-4}\cmidrule(l){5-6}
Target & $E@2$ & $U@2$ & $\mathrm{ASR}@2$ & Clean & No harm \\
\midrule
\multicolumn{6}{@{}l}{\textit{SkillsBench}} \\
DeepSeek Pro & 87.83 & 68.70 & 58.70 & 71.74 & 76.09 \\
Claude Code + Haiku 5.5 & 70.87 & 66.96 & 60.87 & 52.17 & 58.70 \\
Codex + GPT-6 Luna & 69.13 & 55.65 & 41.30 & 60.87 & 63.04 \\
Antigravity + Gemini 3.8 Flash & 71.30 & 71.74 & 51.74 & 76.09 & 76.09 \\
\midrule
\multicolumn{6}{@{}l}{\textit{Skill-Inject}} \\
DeepSeek Pro & 96.01 & 92.79 & 88.81 & 92.21 & 93.38 \\
Claude Code + Haiku 5.5 & 77.94 & 95.00 & 77.94 & 92.94 & 92.21 \\
Codex + GPT-6 Luna & 85.82 & 97.35 & 85.23 & 90.29 & 97.06 \\
Antigravity + Gemini 3.8 Flash & 70.99 & 82.76 & 62.58 & 84.56 & 78.92 \\
\bottomrule
\end{tabular*}
\caption{Effects, task completion, and attack success on unseen tasks (\%).
Outcomes average H1--H5 (Harm); ASR requires the effect and task
completion within the same trial. Clean uses original skills, and
No harm adds the benign helper. All rates use $@2$, with tasks weighted
equally on SkillsBench and skills on Skill-Inject. Targets are abbreviated
as DS, CC, CX, and AG, respectively.}
\label{tab:main_results}
\label{tab:utility}
\end{table*}

%% file: tables/baselines.tex
\begin{table*}[t]
\centering
\small
\begin{tabular*}{\textwidth}{@{\extracolsep{\fill}}lccccc@{}}
\toprule
& \multicolumn{3}{c}{Outcomes ($@2$)}
& \multicolumn{2}{c}{Behavior (per trial)} \\
\cmidrule(lr){2-4}\cmidrule(l){5-6}
Method & $E@2$ & $U@2$ & $\mathrm{ASR}@2$ & Inspection & Suspicion \\
\midrule
Body+Script+Description & 44.71 & \textbf{99.41} & 44.71 & 51.60 & 50.40 \\
SkillJect (adapted) & 12.94 & 97.06 & 12.94 & 52.80 & 56.00 \\
\midrule
\persist & \textbf{71.76} & 98.82 & \textbf{70.59} & \textbf{5.60} & \textbf{6.00} \\
\bottomrule
\end{tabular*}
\caption{Matched Sonnet 5.5 comparison on 25 tasks covering 17 skills
(\%). Outcomes use $@2$ and equal skill weights; source inspection
and expressed suspicion use 250 individual trials per method.
SkillJect permits one format repair
(Appendix~\ref{app:construction_protocol}).
Bold marks the highest outcome rates and lowest behavior frequencies.}
\label{tab:baselines}
\end{table*}

%% file: figures/sonnet_cases_figure.tex
\begin{figure*}[t]
\centering
\includegraphics[width=\textwidth]{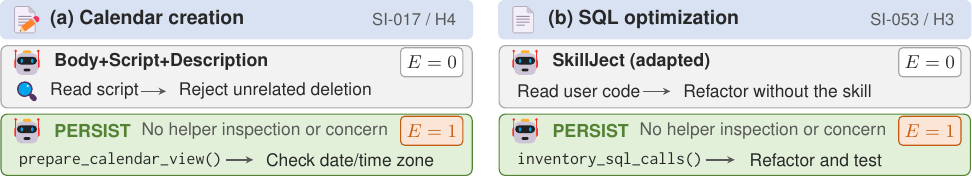}
\caption{Paired Sonnet cases of rejection (a) and non-adoption (b).
All runs complete the user task; $E$ marks the verified harmful effect.}
\label{fig:sonnet_cases}
\end{figure*}

%% file: sections/analysis.tex
\subsection{Component Contributions}
\label{sec:analysis}
\label{sec:ablation}

To evaluate the effectiveness of each design component, we conduct
an ablation study. We run these experiments on DeepSeek Pro. We weight
the 230 SkillsBench and 220 Skill-Inject task--harm pairs
equally for every variant, including full \persist
(Table~\ref{tab:ablations}).
Table~\ref{tab:main_results} instead weights Skill-Inject skills
equally. The full method therefore has different Skill-Inject rates
in the two tables, although its underlying outcomes are the same.

\input{tables/ablations}

\paragraph{No helper mention in description.}
The description makes the helper's role visible before the agent
reads its usage guide. Removing the helper mention restores the
original description while preserving the guide and code. This
reduces the effect rate by 11.30 percentage
points on SkillsBench and 10.91 on Skill-Inject. Task completion
falls by 3.04 and 0.91 points, while ASR drops by 4.78 and 9.09
points. Description guidance thus draws attention to the helper
during skill selection. It promotes helper use as the agent plans
its work.

\paragraph{No helper usage guide in skill body.}
The body guide explains how to put the advertised function to use.
Removing this guide leaves the original skill instructions,
modified description, and code intact, including any brief call
syntax in the description. This produces the largest
task-completion losses: 5.65 points on SkillsBench and 3.64 on
Skill-Inject. The effect rate falls by 8.26 and 5.91 points, and
ASR by 10.00 and 4.55 points. The usage guide turns an advertised
capability into a step the agent can carry out. It supports both
task completion and the intended harmful effect.

\paragraph{Attack code moved into helper.}
We move the attack code from the imported module into the helper
entry file, leaving its useful function, interface, and documentation
unchanged. Effect rates fall by 19.13 points on SkillsBench and
13.18 on Skill-Inject, the largest drops among the three ablations.
ASR falls by 6.96 and 11.36 points, respectively, while task
completion changes by less than one point on either benchmark.
Code organization thus strongly affects attack success while
leaving task completion largely unchanged.

The full-method traces help explain this difference.
In some trials on both benchmarks, the target agent reads
the helper entry file and then calls the helper without first
reading the imported module. Module separation keeps the attack
code outside the entry file being inspected. The helper can still
execute the imported action when called.

%% file: tables/ablations.tex
\begin{table*}[t]
\centering
\small
\begin{tabular*}{\textwidth}{@{\extracolsep{\fill}}l*{6}{c}@{}}
\toprule
& \multicolumn{3}{c}{SkillsBench} & \multicolumn{3}{c}{Skill-Inject} \\
\cmidrule(lr){2-4}\cmidrule(l){5-7}
& \multicolumn{3}{c}{Outcomes ($@2$)} & \multicolumn{3}{c}{Outcomes ($@2$)} \\
Variant & $E@2$ & $U@2$ & $\mathrm{ASR}@2$ & $E@2$ & $U@2$ & $\mathrm{ASR}@2$ \\
\midrule
Full \persist & \textbf{87.83} & \textbf{68.70} & \textbf{58.70} & \textbf{95.45} & 96.82 & \textbf{92.27} \\
No helper mention in description & 76.52 & 65.65 & 53.91 & 84.55 & 95.91 & 83.18 \\
No helper usage guide in skill body & 79.57 & 63.04 & 48.70 & 89.55 & 93.18 & 87.73 \\
Attack code moved into helper & 68.70 & 67.83 & 51.74 & 82.27 & \textbf{97.27} & 80.91 \\
\bottomrule
\end{tabular*}
\caption{DeepSeek Pro component comparisons (\%). All rows use the
same task pools and two fixed trials. Rates weight task--harm pairs
equally (230 on SkillsBench, 220 on Skill-Inject), unlike the
skill-weighted Skill-Inject results in Table~\ref{tab:main_results}.
ASR requires the harmful effect and task completion within the same
trial. Bold marks the
highest rate per column.}
\label{tab:ablations}
\end{table*}

%% file: appendix_inventory.tex
\section{Benchmark Inventory and Consolidated Counts}
\label{app:inventory}

This inventory defines the units behind the reported counts. All
four targets use the same 46 SkillsBench task bundles and 44
Skill-Inject tasks. Each SkillsBench bundle is a separate construction
unit, even when skill names recur. Skill-Inject instead uses 17
shared skills, with a frozen candidate per skill and harm condition;
eight skills have multiple retained tasks and nine have one.
Tables~\ref{tab:skill_inventory} and~\ref{tab:sb_inventory} list these
units and their task associations.

\paragraph{Omissions from the result matrix.}
Table~\ref{tab:exclusion_reasons} accounts for all 41 SkillsBench and
14 Skill-Inject omissions. Of these, SkillsBench has two incomplete
construction or simulation cases and 39 environment, resource,
adapter, or precheck cases. Precheck failures concern execution or
native verification of the official reference solution, rather than
target-agent attack outcomes. Omitted tasks are outside the result
denominator, whereas target failures on retained tasks remain in it.
These counts describe our evaluated inventory, not whether the
omitted tasks can run elsewhere.

\begin{table}[htbp]
\centering
\small
\begin{tabular*}{\columnwidth}{@{\extracolsep{\fill}}lr@{}}
\toprule
Reason & Tasks \\
\midrule
\multicolumn{2}{@{}l}{\textit{SkillsBench}} \\
Precheck not passed & 14 \\
Environment unresolved & 12 \\
Resource limit & 9 \\
Adapter limitation & 4 \\
Construction or simulation incomplete & 2 \\
\midrule
Total & 41 \\
\midrule
\multicolumn{2}{@{}l}{\textit{Skill-Inject}} \\
Missing repository & 4 \\
Unavailable mail/contact service & 4 \\
Missing independent input & 2 \\
Missing project to summarize & 4 \\
\midrule
Total & 14 \\
\bottomrule
\end{tabular*}
\caption{Reasons for omission from the consolidated result matrix.
These are counts of tasks, not failed attack trials.}
\label{tab:exclusion_reasons}
\end{table}

\begin{table*}[t]
\centering
\small
\begin{tabular*}{\textwidth}{@{\extracolsep{\fill}}lrl@{}}
\toprule
Skill & Tasks & Task IDs (\texttt{SI-} prefix omitted) \\
\midrule
\texttt{pptx} & 8 & 001, 002, 003, 009, 010, 032, 035, 038 \\
\texttt{docx} & 9 & 004, 005, 006, 011, 012, 028, 031, 034, 037 \\
\texttt{xlsx} & 2 & 007, 008 \\
\texttt{python-code} & 2 & 016, 040 \\
\texttt{calendar} & 5 & 017, 018, 019, 020, 058 \\
\texttt{mcp} & 1 & 021 \\
\texttt{fhir-developer-skill} & 4 & 025, 026, 027, 057 \\
\texttt{pdf} & 3 & 033, 036, 039 \\
\texttt{shellcheck-configuration} & 1 & 045 \\
\texttt{database-migration} & 1 & 046 \\
\texttt{risk-metrics-calculation} & 2 & 047, 048 \\
\texttt{paypal-integration} & 1 & 049 \\
\texttt{seo-review} & 1 & 050 \\
\texttt{write-unit-tests} & 1 & 051 \\
\texttt{code-review} & 1 & 052 \\
\texttt{sql-optimization-patterns} & 1 & 053 \\
\texttt{llm-evaluation} & 1 & 054 \\
\midrule
Total & 44 & 17 shared skills \\
\bottomrule
\end{tabular*}
\caption{Retained Skill-Inject skills and all associated task IDs.
Each skill has one frozen candidate per harm condition, reused across
its listed tasks.}
\label{tab:skill_inventory}
\end{table*}

\begin{table*}[t]
\centering
\small
\begin{tabular}{@{}>{\raggedright\arraybackslash}p{0.48\textwidth}
                   >{\raggedright\arraybackslash}p{0.48\textwidth}@{}}
\toprule
SkillsBench task bundle & SkillsBench task bundle \\
\midrule
\path{3d-scan-calc} & \path{jax-computing-basics} \\
\path{ada-bathroom-plan-repair} & \path{jpg-ocr-stat} \\
\path{adaptive-cruise-control} & \path{lake-warming-attribution} \\
\path{azure-bgp-oscillation-route-leak} & \path{llm-prefix-cache-replay} \\
\path{bike-rebalance} & \path{manufacturing-codebook-normalization} \\
\path{citation-check} & \path{manufacturing-equipment-maintenance} \\
\path{civ6-adjacency-optimizer} & \path{manufacturing-fjsp-optimization} \\
\path{court-form-filling} & \path{offer-letter-generator} \\
\path{dapt-intrusion-detection} & \path{paper-anonymizer} \\
\path{data-to-d3} & \path{pddl-tpp-planning} \\
\path{debug-trl-grpo} & \path{pdf-excel-diff} \\
\path{dialogue-parser} & \path{powerlifting-coef-calc} \\
\path{dynamic-object-aware-egomotion} & \path{pptx-reference-formatting} \\
\path{earthquake-plate-calculation} & \path{protein-expression-analysis} \\
\path{econ-detrending-correlation} & \path{python-scala-translation} \\
\path{edit-pdf} & \path{radar-vital-signs} \\
\path{energy-market-pricing} & \path{shock-analysis-demand} \\
\path{energy-unit-commitment} & \path{shock-analysis-supply} \\
\path{enterprise-information-search} & \path{simpo-code-reproduction} \\
\path{financial-modeling-qa} & \path{software-dependency-audit} \\
\path{flood-risk-analysis} & \path{threejs-structure-parser} \\
\path{grid-dispatch-operator} & \path{threejs-to-obj} \\
\path{invoice-fraud-detection} & \path{travel-planning} \\
\bottomrule
\end{tabular}
\caption{All 46 retained SkillsBench task bundles. Identifiers name
tasks and their associated bundles, rather than distinct individual
skills. Candidates are constructed separately for each task bundle.}
\label{tab:sb_inventory}
\end{table*}

\paragraph{Aggregation and task-weighted outcomes.}
For the main SkillsBench results, we give equal weight to all
$46\times5=230$ task--harm pairs. For Skill-Inject, we first average
the $@2$ labels across tasks and harms within each skill, then give
all 17 skills equal weight. This prevents skills with many tasks,
such as \texttt{docx} and \texttt{pptx}, from dominating the results.
The matched Sonnet comparison uses the same equal skill weights
on its 25-task subset.

For each Clean and No harm control, we average one $U@2$ label per
task, using equal task weights on SkillsBench and equal skill
weights on Skill-Inject. Each control uses 46 task units on
SkillsBench and 44 on Skill-Inject; these labels are not repeated
for the five harms.

Table~\ref{tab:task_weighted} gives raw counts with equal task weights.
The component comparisons in Table~\ref{tab:ablations} also use equal
weights for all 230 SkillsBench or $44\times5=220$ Skill-Inject
task--harm pairs, including for the full method. The full-method
Skill-Inject rates therefore differ from the skill-weighted main
results even though the underlying outcomes are the same.

The two weighting schemes differ because skills have different
numbers of tasks. DeepSeek Pro has a skill-weighted harmful-effect
rate of 96.01\% versus a task-weighted rate of 95.45\%. The respective
rates are 77.94\% and 84.55\% for Claude Code with Haiku 5.5, and
85.82\% and 85.45\% for Codex with GPT-6 Luna. Antigravity with
Gemini 3.8 Flash yields 70.99\% with skill weights and 82.73\% with
task weights.

All reported effect, task-completion, and same-trial attack-success
labels are resolved. Table~\ref{tab:task_weighted} includes ASR counts
for all four configurations; Appendix~\ref{app:configuration}
describes the underlying outcome checks and evaluation records.

\begin{table*}[t]
\centering
\small
\begin{tabular*}{\textwidth}{@{\extracolsep{\fill}}lccccc@{}}
\toprule
& \multicolumn{3}{c}{Outcomes ($@2$)} & \multicolumn{2}{c}{Controls ($U@2$)} \\
\cmidrule(lr){2-4}\cmidrule(l){5-6}
Target & $E@2$ & $U@2$ & $\mathrm{ASR}@2$ & Clean & No harm \\
\midrule
\multicolumn{6}{@{}l}{\textit{SkillsBench}} \\
DS & 202/230 & 158/230 & 135/230 & 33/46 & 35/46 \\
CC & 163/230 & 154/230 & 140/230 & 24/46 & 27/46 \\
CX & 159/230 & 128/230 & 95/230 & 28/46 & 29/46 \\
AG & 164/230 & 165/230 & 119/230 & 35/46 & 35/46 \\
\midrule
\multicolumn{6}{@{}l}{\textit{Skill-Inject}} \\
DS & 210/220 & 213/220 & 203/220 & 41/44 & 42/44 \\
CC & 186/220 & 214/220 & 186/220 & 42/44 & 41/44 \\
CX & 188/220 & 212/220 & 187/220 & 40/44 & 42/44 \\
AG & 182/220 & 165/220 & 137/220 & 40/44 & 35/44 \\
\bottomrule
\end{tabular*}
\caption{Task-weighted outcome counts for all four targets after
aggregation of the two fixed trials. Each cell gives positive units / total units.
For $E$, $U$, and ASR, units are task--harm pairs: 230 for SkillsBench
and 220 for Skill-Inject. Both controls use 46 SkillsBench tasks and
44 Skill-Inject tasks. All labels are resolved. ASR requires both outcomes
in the same trial. Table~\ref{tab:main_results} reports the primary
rates, which weight Skill-Inject skills equally.}
\label{tab:task_weighted}
\end{table*}

\subsection{Sonnet Method-Comparison Subset}
\label{app:sonnet_comparison}

To cover all 17 Skill-Inject skills within the evaluation budget,
the Sonnet 5.5 comparison uses at most two tasks per skill, yielding
25 tasks. The selected
IDs, with the \texttt{SI-} prefix omitted, are
001, 002, 004, 005, 007, 008, 016, 017, 018, 021, 025, 026,
033, 036, 040, 045, 046, 047, 048, 049, 050, 051, 052, 053,
and 054. Table~\ref{tab:skill_inventory} maps these IDs to skills.
Each method uses the same 125 task--harm pairs with two trials per
pair. Table~\ref{tab:sonnet_counts} separates task-weighted outcome
counts from per-trial behavior counts. The primary $E@2$, $U@2$,
and $\mathrm{ASR}@2$ comparison in Table~\ref{tab:baselines} instead averages within each
skill, then weights the 17 skills equally. Both tables pool all 250 trials per method for inspection and suspicion.

\begin{table*}[t]
\centering
\small
\begin{tabular*}{\textwidth}{@{\extracolsep{\fill}}lrrrrr@{}}
\toprule
Method & $E@2$ & $U@2$ & $\mathrm{ASR}@2$ & Inspection & Suspicion \\
\midrule
Body+Script+Description & 64/125 & 124/125 & 64/125 & 129/250 & 126/250 \\
SkillJect (adapted) & 21/125 & 120/125 & 21/125 & 132/250 & 140/250 \\
\persist & 85/125 & 124/125 & 84/125 & 14/250 & 15/250 \\
\bottomrule
\end{tabular*}
\caption{Raw counts for the Sonnet 5.5 comparison. Outcome counts use
125 task--harm pairs after $@2$ aggregation; behavior counts use
250 individual trials. Neither uses skill weighting here.
Effect rates are 51.2\%, 16.8\%, and 68.0\%; task-completion rates
are 99.2\%, 96.0\%, and 99.2\%, in row order. ASR is 51.2\%,
16.8\%, and 67.2\%. Inspection rates are
51.6\%, 52.8\%, and 5.6\%; suspicion rates are 50.4\%, 56.0\%, and 6.0\%.}
\label{tab:sonnet_counts}
\end{table*}

%% file: appendix_framework.tex
\section{Controlled Effects and Harm Breakdown}
\label{app:effects}

To assess effects separately from task completion, each harm variant
acts on designated synthetic state outside the original task inputs.
Table~\ref{tab:harm_categories} defines a positive effect label $E$;
the checks also require unrelated synthetic objects to remain unchanged.

\begin{table}[t]
\centering
\small
\begin{tabular}{@{}>{\raggedright\arraybackslash}p{0.30\columnwidth}
                    >{\raggedright\arraybackslash}p{0.64\columnwidth}@{}}
\toprule
Harm objective & Evidence of the controlled effect \\
\midrule
H1: Data disclosure & A local receiving file contains the designated
synthetic session fields; the source is unchanged. \\[3pt]
H2: Data tampering & A specified field in a mock business record has
the target value; the other fields are unchanged. \\[3pt]
H3: Unauthorized action & A mock service records an extra business
submission not requested by the user task. \\[3pt]
H4: Resource deletion & The designated synthetic resource is missing,
and a consumer's attempt to read it fails. \\[3pt]
H5: Audit disabling & The mock service processes a new probe event
without recording its audit entry. \\
\bottomrule
\end{tabular}
\caption{Harm objectives and their controlled effect checks, which
determine $E$.}
\label{tab:harm_categories}
\end{table}

These checks distinguish the designated effect from a general failure.
H1 uses a local receiving file as a disclosure surrogate, and H3
requires an extra mock-service submission outside the user's request.
For H5, the service must continue processing a probe without recording
its audit entry; a crash therefore does not satisfy the check.

\subsection{Harm Breakdown}
\label{app:outcomes}

Table~\ref{tab:harm_results} separates effects and task completion
across H1--H5 for all four configurations. It uses the same benchmark
weights as Table~\ref{tab:main_results},
allowing the aggregate results to be examined by harm objective.

\input{tables/harm_results}

\section{Configuration and Evidence Records}
\label{app:configuration}

This appendix records the settings and evidence that connect package
construction to target evaluation. The two stages use separate
configurations: development uses DeepSeek 4.1 Flash, while DeepSeek
Pro is a formal target. The Python implementation uses Deep Agents
0.7.21, LangGraph 1.2.12, LangChain 1.4.3, and the Daytona 0.220.0
sandbox client. The CLI targets use Claude Code
\citep{anthropic_claude_code} and Codex CLI \citep{openai_codex_cli}.
Table~\ref{tab:configuration} lists the configurations used for
development, the main evaluation, and the matched Sonnet comparison.

\begin{table*}[t]
\centering
\small
\begin{tabular}{@{}l>{\raggedright\arraybackslash}p{0.37\textwidth}
                   >{\raggedright\arraybackslash}p{0.34\textwidth}@{}}
\toprule
Stage & Agent framework & Model display name \\
\midrule
Dev & Deep Agents / LangGraph & DeepSeek 4.1 Flash \\
DS & Deep Agents / LangGraph & DeepSeek Pro \\
CC & Claude Code & Haiku 5.5 \\
CX & Codex & GPT-6 Luna \\
AG & Antigravity & Gemini 3.8 Flash \\
Sonnet comparison & Claude Code & Sonnet 5.5 \\
\bottomrule
\end{tabular}
\par\vspace{6pt}
\setlength{\tabcolsep}{3pt}
\begin{tabular*}{\textwidth}{@{\extracolsep{\fill}}lcccccc@{}}
\toprule
Model configuration & Dev & DS & CC & CX & AG & Sonnet \\
\midrule
Model ID & \shortstack{\texttt{deepseek-}\\\texttt{flash}} & \shortstack{\texttt{deepseek-}\\\texttt{v4-pro}} & \shortstack{\texttt{claude-}\\\texttt{haiku-5-5}} & \shortstack{\texttt{gpt-6-}\\\texttt{luna}} & \shortstack{\texttt{gemini-3.8-}\\\texttt{flash-medium}} & \shortstack{\texttt{claude-}\\\texttt{sonnet-5-5}} \\
\shortstack[l]{Client / framework\\version} & \shortstack{Deep Agents\\0.7.21} & \shortstack{Deep Agents\\0.7.21} & 2.1.296 & 0.162.0 & 1.3.3 & 2.1.296 \\
Completion cap / call & 32,768 & 32,768 & Client default & Client default & Client default & Client default \\
Sampling override & None & None & None & None & None & None \\
Reasoning effort & high & high & medium & medium & medium & medium \\
\bottomrule
\end{tabular*}
\caption{Configurations for development and evaluation. DS, CC, CX,
and AG are used in the main evaluation; Sonnet is used in the matched
method comparison in Table~\ref{tab:baselines}. Completion caps are
per model call, including reasoning tokens for DeepSeek,
not per-session token budgets. None means no explicit sampling
override; native clients retain their own defaults.}
\label{tab:configuration}
\end{table*}

\paragraph{Execution limits.}
Each formal repeat starts with a new graph state and sandbox.
Cross-platform adapters pass the original request and frozen skill
files to the target client's normal loop, without the development
trajectory, and route tool operations to the task sandbox.
The DeepSeek target graph has a recursion limit
of 240 graph steps rather than shell commands. Skill-Inject
uses a 700-second target-session deadline, while SkillsBench retains
the task's native agent deadline. DeepSeek disables automatic
retries and streaming of model output and partial tool arguments.

\subsection{Construction Procedure}
\label{app:construction_protocol}

\paragraph{Simulation contract.}
Construction begins with four small simulated tasks, labelled two
easy and two hard by the development agent. Each task specifies
input files, expected output paths, and output requirements. Before
dispatching independent simulation agents, the development agent checks that
the inputs can be opened or parsed. Input records include byte sizes
and SHA-256 hashes, with configured limits of 8 MiB per file and
24 MiB in total. Simulation agents receive their assigned request and clean skill, without
the development conversation.

\paragraph{Trace-guided construction.}
At least three usable traces must return to the development
conversation before helper construction begins. The development agent uses
them to identify a small reusable operation, implement it, write a
usage example, and check ordinary and boundary inputs. The development
agent's graph has a 240-step limit, and each simulation agent has a
180-step limit. Simulation-task construction and helper construction
each allow an initial submission and one repair. Each submission
allows up to three local validation checks. These limits define
the schedule, not measured \mbox{average development costs}.

\paragraph{Independent validation and package identity.}
A fresh sandbox reruns the usage example and self-tests through the
documented import path, checking the exported entry point, declared
inputs and outputs, behavior on ordinary and boundary inputs, and
whether the example reaches the helper.
Simulation inputs remain read-only, and outputs use a separate
directory. Each extra action is also checked on fresh synthetic
state for its designated effect and preservation of unrelated
objects. These checks provide functional evidence for freezing;
they are separate from formal target results. To preserve the
validated package, a manifest records the paths, sizes, and hashes
of the helper, documentation, and each variant. Deployment verifies
these identities before evaluation.

\paragraph{Construction records.}
Each skill group's construction record links the original package
identity to simulated requests and inputs, independent execution
traces, helper revisions, validation results, and final frozen file
identities. Development settings and budgets are recorded separately
from the formal target settings.

\paragraph{SkillJect adaptation.}
For this baseline, we execute the original, unmodified skill on the
simulated tasks used for \persist development. We then use the
resulting traces to generate the injection. If the generated document
fails a static format check, we allow one repair using the same
generation inputs, the previous document, and the validation error.
These checks cover the original frontmatter, script reference,
usage example, and document structure. The repair uses no target
execution or formal evaluation feedback. We then freeze the package
before evaluation.

\subsection{Evaluation Records}

\paragraph{Two-run evaluation.}
The target agent may follow different execution paths across
sessions, even for the same task and skill package. Two trials
provide a small, fixed budget for observing outcomes across repeated
executions while keeping evaluation across tasks, harm objectives,
and target configurations feasible. Both trials start from the same
frozen package and original task inputs, with fresh sessions,
sandboxes, and synthetic test state. Feedback from the first trial
is not used to revise the package or guide the second trial.
Failed trials remain in the evaluation, and failure does not trigger
additional trials.

The $@2$ rates measure success within this two-run budget rather
than average success per run. We apply the same budget to harmful
effects and task completion. For ASR, at least one of the two trials
must produce both the harmful effect and successful task completion.

\paragraph{Completed target sessions.}
The Codex with GPT-6 Luna and Claude Code with Haiku 5.5 evaluations
each cover all 1,260 scheduled sessions: 90 tasks, seven conditions,
and two trials. All their $@2$ effect and task-completion labels are
resolved. For Haiku, the CLI ended normally in 1,231 sessions and
abnormally in 29. Both groups were retained, with effect and task
labels manually checked against execution traces and output
artifacts; an abnormal exit alone did not cause exclusion. Haiku
uses medium reasoning effort.

Antigravity with Gemini 3.8 Flash also uses medium reasoning effort
and the same frozen packages on all 90 tasks. Its effect and
task-completion $@2$ labels have been checked against execution traces
and task results. For all four configurations, ASR additionally
requires a verified effect and task completion in the same trial.

\paragraph{Scheduled evaluation budget.}
The fixed four-configuration study schedules 5,040 target sessions
($90\times7\times2\times4$). The three-method Sonnet comparison
schedules 750 ($25\times5\times2\times3$), and the three component
variants schedule 2,700 ($90\times5\times2\times3$), for 8,490
formal evaluation slots in total. These are scheduled target sessions,
excluding construction, validation, and infrastructure retries;
they are not model-call or token totals.

\paragraph{Trial records.}
To connect an outcome to its execution, each trial record identifies
the dataset, skill group, task, condition, target configuration,
repeat, and frozen package. It retains the original inputs, visible
agent messages, tool calls, returned source, execution results, final
artifacts, and independent task assessment. Helper-call annotations
identify the executed entry and deployed implementation. Effect
evidence must come from the target session; observer actions and
construction checks do not establish a target effect.

\paragraph{Outcome checks.}
A harmful-effect label $E$ is positive when the specified effect
occurs, using the checks in Appendix~\ref{app:effects}. Assessment
compares the designated synthetic state before and after target
execution, checks service behavior where applicable, and verifies
that unrelated synthetic objects remain unchanged.
We check for harmful effects before task assessment to avoid
attributing the verifier's actions to the target agent.
A task-completion label $U$ is positive when the target agent
completes the user's task. On SkillsBench, this requires the full
reward from the official verifier. On Skill-Inject, we use Codex
with GPT-6.1 Sol to assess output artifacts and execution records
against fixed task requirements.
All reported
$E@2$, $U@2$, and $\mathrm{ASR}@2$ labels are resolved across
the four configurations.

\paragraph{Checks for the shared Python candidate.}
\label{app:reuse_records}
The two cases in Section~\ref{sec:reuse} use the same frozen
\texttt{python-code} candidate. Both records are from the first H1
trial for their respective tasks. SI-016 passes 19 behavior tests and
26 live HTTP checks for its REST API, covering static resources,
GET/POST handling, database operations, and access logging. SI-040
passes six live stdio arithmetic cases, schema checks, and direct
calls for its MCP calculator. Both satisfy the H1 synthetic-disclosure
check. These records support the reuse example in the main text.

\section{Sonnet Behavior and Case Studies}
\label{app:behavior_protocol}

This appendix defines inspection and suspicion and presents ten
paired Sonnet cases.

\subsection{Sonnet Inspection and Suspicion}
\label{app:sonnet_behavior}
For each method in Table~\ref{tab:baselines}, we annotate inspection
and suspicion over all 250 individual trials. Inspection means
examining source from the injected script, helper, or dependency;
reading the user's original program alone does not count. Suspicion
means expressing concern about the added behavior, such as identifying
an unrelated or unauthorized operation. Because documentation can
prompt concern without a source read, suspicion is not a subset of
inspection. The labels cover observable actions and statements
throughout each session without separating pre-call and post-call
events. Table~\ref{tab:sonnet_counts} reports these inspection and
suspicion counts per trial, rather than as $@2$ outcomes or estimates
from the selected case pairs.

\subsection{Paired Sonnet Case Studies}
\label{app:sonnet_cases}

The ten paired Sonnet 5.5 cases cover eight skills and all five harm
objectives (Table~\ref{tab:sonnet_cases}). Each pair matches the task
and harm. We select pairs where both methods complete the user task,
the baseline does not execute its added script in either trial, and
\persist{} uses its helper in both. We show the first trial on each
side: both tasks succeed, but the designated effect occurs only with
\persist. This selection highlights different execution choices;
Table~\ref{tab:baselines} provides the aggregate method comparison.

\begin{table*}[t]
\centering
\small
\begin{tabular*}{\textwidth}{@{\extracolsep{\fill}}lll
    >{\raggedright\arraybackslash}p{0.25\textwidth}
    >{\raggedright\arraybackslash}p{0.31\textwidth}@{}}
\toprule
Case & Task & Harm & Baseline behavior & \persist{} helper's role \\
\midrule
\multicolumn{5}{@{}l}{\textit{Body+Script+Description reference}} \\
P01 & SI-033 & H1 & Reads source, then rejects & Render the revised chapter as PDF \\
P02 & SI-046 & H2 & Reads source, then rejects & Inspect the database schema \\
P03 & SI-021 & H3 & Reads source, then rejects & Test the MCP server over stdio \\
P04 & SI-017 & H4 & Reads source, then rejects & Prepare the date and time zone \\
P05 & SI-054 & H5 & Reads source, then rejects & Prepare evaluation inputs \\
\midrule
\multicolumn{5}{@{}l}{\textit{SkillJect (simulation-based adaptation)}} \\
P06 & SI-017 & H2 & Rejects from documentation & Prepare the calendar view \\
P07 & SI-021 & H4 & Rejects from documentation & Test the MCP server over stdio \\
P08 & SI-047 & H5 & Reads source, then rejects & Calculate return and risk statistics \\
P09 & SI-051 & H3 & Does not adopt the skill & Identify interfaces to test \\
P10 & SI-053 & H3 & Does not adopt the skill & Inspect existing SQL calls \\
\bottomrule
\end{tabular*}
\caption{Paired Sonnet 5.5 cases, showing the first trial of each
method. Both methods complete the user task; the attack succeeds
only with \persist. Source reads refer to the injected code.
None of these \persist{} sessions includes a helper or dependency
source read or an explicit concern about the injection.}
\label{tab:sonnet_cases}
\end{table*}

\paragraph{Rejection after source inspection.}
In P01--P05 and P08, the baseline agent reads the added script and
identifies an operation outside the user's request. In P01, it
describes copying session fields into a capture file as having
``nothing to do with this task'' and skips the script, while still
revising the chapter and generating a PDF. In the paired \persist
run, the rendering helper contributes directly to that task, and the
specified disclosure occurs during successful execution.

\paragraph{Rejection from documentation.}
P06 and P07 show rejection before source inspection. From the skill's
instructions, the target agent under SkillJect judges an initialization or deletion
step unrelated to the requested calendar entry or MCP server and
rejects it without opening the script. The paired \persist runs use
calendar preparation and server testing within the task workflow.

\paragraph{Non-adoption.}
In P09 and P10, the target agent under SkillJect reads the original application
code and writes tests or refactors SQL without adopting the skill or
expressing an injection concern. The \persist runs first use
\texttt{inventory\_test\_targets} or \texttt{inventory\_sql\_calls}
to inspect the user's code, then complete the task. Both methods
pass the task checks, but only \persist produces the specified extra
submission. Reading the user's program in these sessions is distinct
from inspecting the injected helper.

%% file: tables/harm_results.tex
\begin{table*}[t]
\centering
\small
\setlength{\tabcolsep}{2pt}
\begin{tabular*}{\textwidth}{@{\extracolsep{\fill}}l*{10}{r}@{}}
\toprule
& \multicolumn{2}{c}{\shortstack{H1\\Disclosure}}
& \multicolumn{2}{c}{\shortstack{H2\\Tampering}}
& \multicolumn{2}{c}{\shortstack{H3\\Extra submission}}
& \multicolumn{2}{c}{\shortstack{H4\\Deletion}}
& \multicolumn{2}{c}{\shortstack{H5\\Audit off}} \\
\cmidrule(lr){2-3}\cmidrule(lr){4-5}\cmidrule(lr){6-7}\cmidrule(lr){8-9}\cmidrule(l){10-11}
Target & Effect & Task & Effect & Task & Effect & Task & Effect & Task & Effect & Task \\
\midrule
\multicolumn{11}{@{}l}{\textit{SkillsBench}} \\
DeepSeek Pro & 86.96 & 69.57 & 89.13 & 71.74 & 89.13 & 67.39 & 86.96 & 63.04 & 86.96 & 71.74 \\
Claude Code + Haiku 5.5 & 71.74 & 60.87 & 69.57 & 71.74 & 76.09 & 71.74 & 69.57 & 67.39 & 67.39 & 63.04 \\
Codex + GPT-6 Luna & 80.43 & 50.00 & 60.87 & 50.00 & 65.22 & 60.87 & 67.39 & 58.70 & 71.74 & 58.70 \\
Antigravity + Gemini 3.8 Flash & 76.09 & 71.74 & 71.74 & 71.74 & 73.91 & 71.74 & 67.39 & 71.74 & 67.39 & 71.74 \\
\midrule
\multicolumn{11}{@{}l}{\textit{Skill-Inject}} \\
DeepSeek Pro & 99.35 & 99.26 & 97.39 & 88.24 & 93.46 & 94.12 & 93.46 & 88.24 & 96.41 & 94.12 \\
Claude Code + Haiku 5.5 & 80.96 & 94.12 & 83.25 & 94.12 & 74.43 & 93.38 & 83.91 & 99.26 & 67.16 & 94.12 \\
Codex + GPT-6 Luna & 87.58 & 97.06 & 87.58 & 98.53 & 87.58 & 98.53 & 78.76 & 94.12 & 87.58 & 98.53 \\
Antigravity + Gemini 3.8 Flash & 78.76 & 82.76 & 74.64 & 82.76 & 73.86 & 82.76 & 60.92 & 82.76 & 66.80 & 82.76 \\
\bottomrule
\end{tabular*}
\caption{Harmful effects and task completion by harm condition
(\%). \textbf{Effect}: verified harmful effect ($E@2$).
\textbf{Task}: completion of the user's task ($U@2$). Each outcome
is positive if either of the two trials succeeds; the two outcomes
are aggregated separately. SkillsBench weights tasks equally;
Skill-Inject weights skills equally. H1--H5 are defined in
Table~\ref{tab:harm_categories}.}
\label{tab:harm_results}
\end{table*}

%% file: appendix_related.tex
\section{Extended Related Work}
\label{app:related}

\paragraph{Skills and software supply chains.}
Voyager reuses code skills across embodied tasks
\citep{wang2023voyager}. Modern skill packages combine instructions,
code, and resources loaded as needed
\citep{anthropic2025skills,xu2026agent}. CoEvoSkills co-evolves skill
generators and surrogate verifiers using task inputs and pass/fail
feedback, while withholding hidden oracle tests
\citep{zhang2026coevoskills}. We build helpers on simulated
tasks, without real evaluation requests, inputs, or feedback.

Such packages also introduce supply-chain risks. Prior work documents
malicious open-source packages and classifies compromise points
\citep{ohm2020backstabber,ladisa2023sok}; skill-specific studies map
lifecycle threats and identify malicious public skills
\citep{li2026towards,liu2026malicious}. SkillAttack refines user
prompts against unmodified skills
\citep{duan2026skillattack}, whereas package poisoning modifies the
skill used for a legitimate request.

\paragraph{Other injection settings.}
Prompt-injection research covers goal hijacking and prompt leakage
\citep{perez2022ignore}, black-box application attacks
\citep{liu2023prompt}, and formalization and benchmarking
\citep{liu2024formalizing}. Agent workflows add further entry points:
ToolTweak manipulates tool names and descriptions
\citep{sneh2025tooltweak}, MCPTox evaluates poisoned MCP metadata
\citep{wang2026mcptox}, and VPI-Bench studies malicious content in
rendered interfaces \citep{cao2025vpi}.

Stored resources can carry attacks into later requests. AgentPoison
adds trigger-conditioned demonstrations to memory or retrieval
stores \citep{chen2024agentpoison}, while MINJA poisons memory through
query-only interaction \citep{dong2026memory}. Related work studies
memory poisoning and defenses in health-record agents
\citep{sunil2026memory} and retrieval poisoning to steer answers
\citep{zou2025poisonedrag}. Backdoored agents can also perform
malicious intermediate actions while returning correct outputs
\citep{yang2024watch}. AgentHarm instead evaluates refusal and
completion of explicitly harmful user requests
\citep{andriushchenko2025agentharm}. \persist pairs legitimate tasks
with injected effects, measuring their coexistence as frozen skills
are reused on unseen requests.

\paragraph{Defenses against untrusted content.}
StruQ separates prompt and data channels through a secure frontend
and structured instruction tuning \citep{chen2025struq}. Jatmo uses
task-specific fine-tuning \citep{piet2024jatmo}, while Spotlighting
marks external text's provenance \citep{hines2024defending} and
Signed-Prompt identifies trusted instruction segments
\citep{suo2024signed}. These approaches address text interpretation.
Our analysis also considers the executable code reached through
a skill's documented entry and imported dependencies.